\documentclass[pre,floatfix,twocolumn, 10pt]{revtex4} 
\usepackage{graphicx}
\usepackage{natbib}
\usepackage{dcolumn}
\usepackage{bm}
\usepackage{color}
\usepackage[colorlinks=true, linkcolor=blue, citecolor=blue]{hyperref}
\usepackage{subfigure}
\usepackage{graphicx,psfrag,xspace}
\usepackage{color}
\usepackage{amssymb,amsfonts,amsmath}
\usepackage{setspace}

\begin{document}

%\author{E. E. Ferrero} 
%\affiliation{Instituto de Nanociencia y Nanotecnolog\'{\i}a, CNEA--CONICET, 
%Centro At\'omico Bariloche, (R8402AGP) San Carlos de Bariloche, R\'{\i}o Negro, Argentina.}

\author{E. A. Jagla} 
\affiliation{Centro At\'omico Bariloche, Instituto Balseiro, 
Comisi\'on Nacional de Energ\'ia At\'omica, CNEA, CONICET, UNCUYO,\\
Av.~E.~Bustillo 9500 (R8402AGP) San Carlos de Bariloche, R\'io Negro, Argentina}

\title{From shear bands to earthquakes in a model granular material with contact aging}

\begin{abstract} 

We perform molecular dynamics simulations of homogeneous athermal systems of poly-disperse soft discs under shear. 
For purely repulsive interactions between particles, and under a confining external pressure,  a monotonous flow curve (strain rate vs. stress) starting at a critical yield stress is obtained, with  deformation distributing uniformly in the system, on average. Then we add a short range attractive contribution to the interaction potential that increases its intensity as particles remain in contact for a progressively longer time, mimicking an aging effect into the system.
In this case the flow curve acquires a reentrant behavior, namely, a region of negative slope. 
Within this region the deformation is seen to localize in a shear band with a well defined width that decreases as the global strain rate does. At very low strain rates
the shear band becomes very thin and deformation acquires a 
prominent stick-slip behavior. This regime can be described as the system possessing a fault in which deformation occurs with the phenomenology that characterizes earthquakes. In this way the system we are analyzing connects a regime of uniform deformation at large strain rates, a localized deformation regime in the form of shear bands at intermediate stain rates, and seismic phenomena at very low strain rate. The unifying ingredient of this phenomenology is the existence of a reentrant flow curve, originated in the aging mechanisms present in the model.

\end{abstract}

\maketitle

\section{Introduction}

Yield stress materials (YSMs) is the generic name for a broad range of materials
that display elastic response under applied shear below some threshold, and fluid-like behavior when this threshold is
exceeded\cite{coussot,bonn,ovarlez,ferrero_rmp}. 
The threshold value of the applied stress defines the yield stress $\sigma_c$ of the material.
Structurally, YSMs are usually amorphous systems, in many cases in the category of granular systems.
In this last case
one characteristic parameter of the material is the typical size of its ``grains". This size can range from atomic size to macroscopic.
We focus in cases in which the elementary grain size is sufficiently large that the effect of thermal fluctuations is negligible. These materials can be safely considered as athermal.

One of the main characteristics of a YSM is its {\em flow curve} which, 
for a simple shear geometry, is the dependence of the shear stress $\sigma$ on the strain rate $\dot\gamma$ (note however that in different experiments, either $\sigma$ or $\dot\gamma$ can be the externally controlled variable).
We consider here only cases in which the flow curve is well defined, and independent on the history of the sample, thus leaving aside materials displaying other phenomena such as tixotrophy\cite{barnes,tixo}.

The simplest form of a flow curve is one in which $\dot\gamma=0$ if $\sigma<\sigma_c$, with $\dot\gamma$ increasing continuously as $\sigma$ is increased from $\sigma_c$. When this occurs, the value of $\dot\gamma$ is typically not only the average deformation rate of the whole sample, but also the time average of the deformation at any single point in the sample. In fact, considering the sample as a stack of layers parallel to the direction of applied shear, mechanical equilibrium requires that stress is the same on each layer, and therefore the local strain rate must be everywhere equal to the only value compatible with the local stress. 
Experimental measurements agree with this picture (in particular for systems of repulsive particles which typically have this kind of flow curve), where it is found that deformation is uniform if observed over sufficiently long times\cite{ovarlez2010,divoux,becu}.

There are cases however in which the deformation of YSMs is known to be non-homogeneous. 
In these cases, under a homogeneous applied shear, the material deformation may be typically localized in a region known as a shear band. For instance, failure by shear banding is well known to occur in metallic glasses\cite{mg1,mg2,mg3,mg4}.
However, the case of metallic glasses is an example where the localized deformation originates in the particular form of sample preparation, and in this way it can be classified as a non-stationary phenomenon.

On the theoretical side, it has been realized that there are cases in which non-homogeneous deformation can occur as a permanent phenomenon in a YSM. 
The simplest possibility to have such a situation is if the
flow curve of the system (assumed {\em a priori} to describe a homogeneous situation) has a reentrant part, namely a region with $d\sigma/d\dot\gamma <0$. 
In this case, similarly to what occurs in the coexistence region of first order phase transitions, under an applied $\sigma$ in the reentrant region, the system may separate in two parts, flowing with the two different values of $\dot\gamma$ compatible with the applied $\sigma$. A number of different models have been proposed that generate a reentrant flow curve, and thus the appearance of shear bands \cite{jagla2007,fielding,sollich,coussot2010,mansard,martens,vdb}, which typically introduce an internal time scale in the system associated to some kind of aging that competes with the time scale imposed by the applied $\dot\gamma$.
Qualitatively, under such a relaxation process, some parts of the system can remain stuck, in a very relaxed configuration, while other regions yield, continuously refreshing their structure, and therefore not being able to relax. The mechanical and geometrical constrains onto the system require that the yielding part has the structure of a layer parallel to the applied shear stress, therefore defining a shear band into the system.

A slightly different possibility for the appearance of permanent shear bands is the case in which the dynamics of the system is such that two different branches of the flow curve coexist. One normal (i.e., non-reentrant) flow curve with some critical stress $\sigma_{c1}$, and a frozen branch with $\dot\gamma=0 $ between $\sigma=0$ and $\sigma=\sigma_{c2}>\sigma_{c1}$. In this case, if $\sigma_{c1}<\sigma <\sigma_{c2}$, there are effectively two possible values of $\dot\gamma$, and the phenomenological situation is in fact very similar to the case of a reentrant flow curve.
This possibility was first suggested to occur in systems of Lennard Jones particles \cite{varnik,berthier}, and then it was seen to be rather generic of systems with attractive particles (specially in the case in which this attraction is very short ranged) \cite{irani,chauduri}. The appearance of shear bands in systems of attractive particles has received experimental confirmation\cite{becu,ragouilliaux,fall,paredes}.

%Although attraction between particles can produce shear bands due to the appearance of two branches of the flow curve, we emphasize that this mechanism is different from the one described in the previous paragraph where a single, reentrant flow curve appears due to aging mechanisms in the sample. 
We want to consider here a sort of combination between the two mechanisms for the appearance of shear band mentioned in the two previous paragraphs. We will consider attraction between particles, but allow this attraction to age in time, becoming stronger as the particles hold together for longer times. In considering the physical basis for 
this possibility, notice that the particles we refer to, do not need to be microscopic, they may well represent  macroscopic grains of a  granular material, and here the possibility of aging at the contacts emerges very naturally. In fact, contact aging refers to the (typically logarithmic)
 strengthening of the interaction with hold time between two bodies in contact. Although the effect is clear and well known at the largest scale given by earthquakes (where it is phenomenologically incorporated in the rate-and-state equations of friction\cite{dietrich,ruina,scholz}), it has also been clearly observed in mesoscopic contacts with the aid of Atomic Force Microscope. \cite{li,feldmann,vorholzer1,vorholzer2} 

The aim of the present work is to incorporate in molecular dynamics simulations the possibility of aging through a time dependent attraction, which becomes stronger as the time-into-contact between the particles increases. 
We start by considering a model of soft repulsive particles under a confining external pressure. 
In this condition the model is jammed, with a finite yield stress $\sigma_c$, and a monotonously increasing flow curve as $\sigma>\sigma_c$. On this basic model, we add attraction between the particles. However the attraction potential does not depend only on distance between particles, but also on time, becoming stronger as the hold time in
contact configuration is increased, therefore generating aging in the system. The time scale set by the external driving competes with the time scale of the attraction build up, producing a flow curve in the system with a reentrance, and stationary shear bands when the system is driven at a constant strain rate. After characterizing this regime, we focus on what happens when 
$\dot\gamma$ becomes very small. We observe that the shear band becomes very thin, and that the sliding acquires a strong stick-slip character, that we associate to the emergence of earthquake-like avalanches in the system. To 
reinforce this idea, we show the existence of  aftershocks, one of the hallmarks of seismic phenomena. The crossover between shear bands at relatively large strain rates, and earthquake phenomenology at very low strain rate is one of the main results of our investigation.

In the next Section we introduce the aging mechanism in one of the simplest mean field examples of a yield stress system, namely, the Prandtl-Tomlinson model of friction.  Then Section III describes the main results of the paper, corresponding to a system of soft disks with an aging attraction,  presenting the phenomenology of shear bands, and its transition to earthquake phenomena at very los strain rates. Section IV contains a summary and the conclusions.

\section{The Prandtl-Tomlinson model with aging}

One of the prototypical models that has been largely used to understand qualitatively the elemental processes
at play during friction is the Prandtl-Tomlinson (PT) model.\cite{pt1,pt2,pt3} In its standard version it describes the friction between solid bodies in contact, but it can easily be considered as describing the possibility of slip between two adjacent layers of a bulk three dimensional sample, and in this way it is of fundamental interest to us.

\begin{figure}
\includegraphics[width=7cm,clip=true]{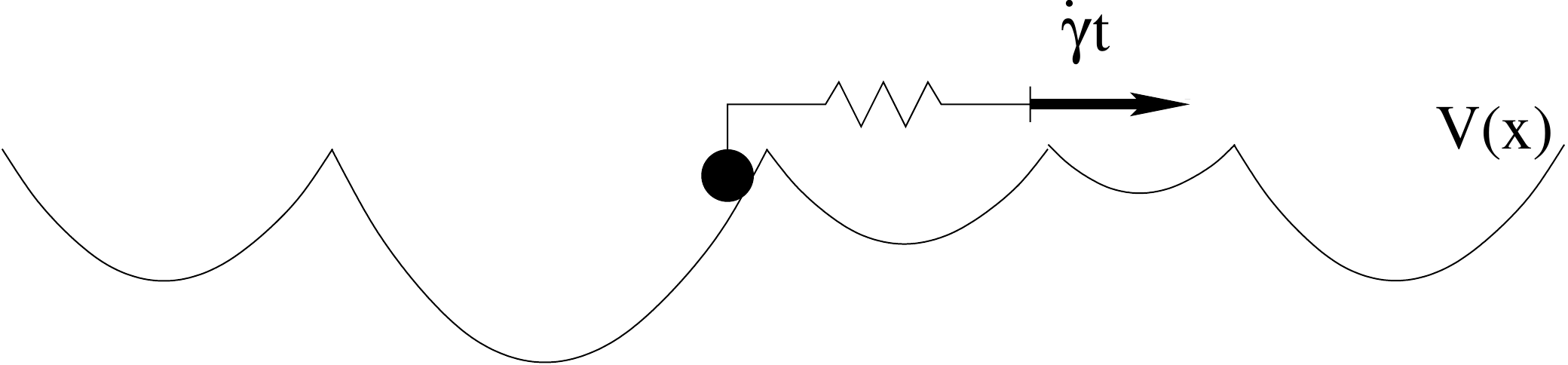}
\caption{The basic scheme of the Prandtl-Tomlinson model, on a piece-wise parabolic potential.}
\label{pt}
\end{figure}

The two layers are descried in the PT model by a mass, and a potential on which this mass moves (see Fig. \ref{pt}). It qualitatively represents the interlocking potential between two adjacent layers of material, or two solid bodies sliding pass each other. The potential is fixed, and the mass is driven by a spring, moved from the other tip at a constant velocity $\dot\gamma$ that represents the strain rate applied to the material, while the spring $k_0$ represents the elasticity of the material. The potential is expected to have many minima in different spatial positions, represented the corrugation of the interaction between the two layers. Beyond this fact, its precise form it is not crucial. For concreteness we construct the potential in the following way. It is a concatenation of parabolic pieces. The width $\Delta$ of each piece is chosen from a flat probability distribution extending between $\Delta_{min}$ and $\Delta_{max}$. The dynamics of the mass on top of this potential is of the overdamped form, and is given by the following equation 
\begin{equation}
\dot x=-\frac{dV}{dx}+(\dot \gamma t-x)k_0
\label{pt_eq}
\end{equation}
The instantaneous stress $\sigma$ in the system is measured from the stretching of the spring, namely
\begin{equation}
\sigma=(\dot \gamma t-x)k_0
\label{sigma}
\end{equation}

\begin{figure}
\includegraphics[width=9cm,clip=true]{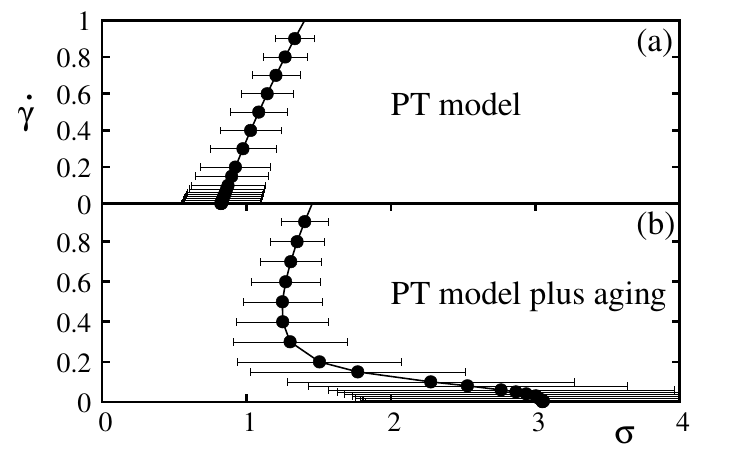}
\caption{Results for strain rate $\dot\gamma$ vs stress $\sigma$ in the PT model, using $k_0=0.3$, $\Delta_{min}=0.5$, $\Delta_{max}=1.5$. (a) Results in the standard 
model. (b) Results when aging is added (we use $f_{max}=4$, $\tau=10$. Note that the simulations are done by fixing $\dot\gamma$ and measuring $\sigma$).}
\label{pt_results}
\end{figure}

The simulation of Eq. \ref{pt_eq}, with definition \ref{sigma}
produces the plot shown in Fig. \ref{pt_results}(a). This is the flow curve of the system. At $\dot\gamma =0^+$
the stress attains a value $\sigma_c$, which defines the critical stress in the system. As $\dot\gamma$ is increased, 
$\sigma$ increases monotonously. We see that this increase is linear, namely 
$\dot\gamma=C(\sigma-\sigma_c)$. This linear dependence is related to the corners joining adjacent parabolic pieces of the potential. If we had used a continuous function for the potential (for instance $\sim \cos(x)$), the dependence we had obtained is $\dot\gamma=C(\sigma-\sigma_c)^{3/2}$. In addition to plot the average value of $\sigma$ along a simulation, we show also in Fig. \ref{pt_results}(a) the deviation $\delta \sigma=\sqrt{\overline{\sigma^2}-\overline{\sigma}^2}$ in the form of a bar between $\sigma\pm\delta\sigma$. Note that this is not the error in the estimation of the mean value, but a 
measurement of the natural fluctuations of a dynamical process. 
%To complete the results, we see in Fig. \ref{pt_results2} the temporal evolution of the instantaneous value of $\sigma$ at a few different values of $\dot\gamma$.

The central goal of the PT model is to explain the finite value of $\sigma$ in the limit $\dot\gamma\to 0$. Note that this occurs despite the underlying dissipation mechanism is viscous (the velocity term in Eq. (\ref{pt_eq})). The
reason is in the fact that the velocity of the mass is finite when jumping from one potential well to the next, and this generates a finite dissipation even if $\dot\gamma$ is vanishingly small.

Up to here the classical results on the PT model. Now we are going to include in the model an {\em aging} mechanism (see \cite{carpick} for a recent approach along the same line). The qualitative motivation is the following.
The potential in the PT model is a sort of stickiness between the atoms lying on two adjacent layers of material. The idea is to propose that this stickiness becomes stronger as the system remains in the same configuration for a longer time. 
In concrete, we take the interaction potential of a given parabolic well, and increase its strength as a function of the time that the mass has spent within the current potential well. Namely, the potential becomes also a function of time in addition to space, of the form

\begin{equation}
V(x,t)=(x-x_0)^2G(t-t_0)
\end{equation}
Here $x_0$ is the mean position of the potential well, and $G(t-t_0)$ is the ``aging function" that takes the value 1
at $t=t_0$ (where $t_0$ is the time at which the mass entered the current well) and is an increasing function of $t$. Qualitatively, this models a potential that becomes deeper as the mass spends more and more time in it. The exact form in which $G$ increases with time is not crucial. We will take for concreteness an exponential saturation of the $G$ function from the initial value 1 to a final value $G_{max}$, with a time constant $\tau$, namely

\begin{equation}
G(t-t_0)= (1-G_{max})\exp\left (-(t-t_0)/\tau \right )+G_{max}
\end{equation}

\begin{figure}
\includegraphics[width=9cm,clip=true]{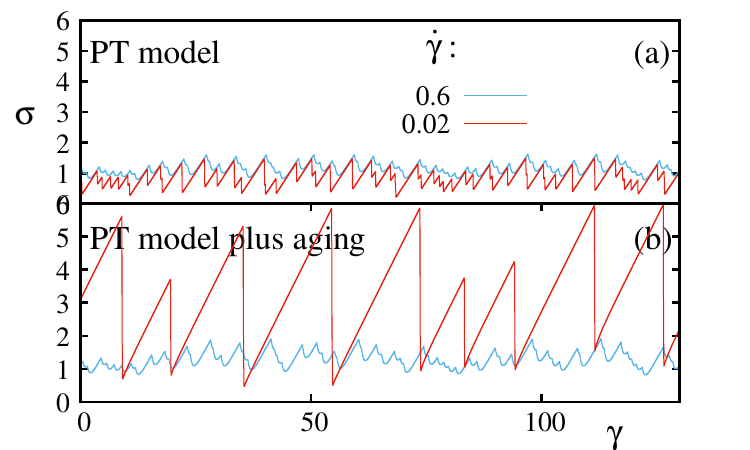}
\caption{Temporal evolution of stress at two different values of strain rate, in the standard PT model(a) and when aging is included (b). The strong stick-slip effect at low strain rate is clearly visible in the second case.}
\label{trazas}
\end{figure}

The simulation of the model with the time dependent potential can be done in the same way than in the previous case.
The results obtained for $\sigma$ and $\delta \sigma$ are presented in Fig. \ref{pt_results}(b). The first remarkable point in these results is the reentrant form of the flow curve at low values of $\dot\gamma$, originated in a progressive increase of $\sigma$ as $\dot\gamma$ is reduced, as compared with the case without aging.
The origin of this behavior is easy to understand. If $\dot\gamma$ is large, the mass spends a short time in each potential well, the potential does not have enough time to age, and the measured $\sigma$ value is similar to the case without aging. As $\dot\gamma$ is reduced the mass spends more time in each potential well, the force necessary to escape from it increases, and so does the average stress $\sigma$. 
The increase of the maximum force to escape a given potential well due to aging has effect in the temporal fluctuations of $\sigma$. In fact, it is seen in Fig. \ref{pt_results}(b) that this fluctuation increases strongly in the reentrant region of the flow curve. Figure \ref{trazas} displays the temporal evolution of $\sigma$ at two different values of $\dot\gamma$, with and without aging. We see that aging strongly increases the fluctuation of $\sigma$ at low values of $\dot\gamma$. In fact, in this case the slow driving allows the potential to age, and the value of stress needed to jump from one potential well to the following increases roughly from a value $\Delta_{max}$ to the aged value $\Delta_{max}G_{max}$.

\section{Simulations of a system of soft disks with contact aging}

The behavior of the PT model with aging is a reference point for analyzing the behavior of a full system under shear. Conceptually we can consider a bulk system as a collection of layers that are forced to shear, on average, at an imposed strain rate $\dot\gamma$. However, whether this deformation distributes uniformly in all layers or not will depend on the characteristics of the flow curve. A mechanical equilibrium consideration indicates that the stress must be the same in all layers. If the underlying flow curve of each layer is monotonous, the strain rate of each layer will be equal to the externally imposed value. However, if the flow curve is reentrant, the uniform strain rate configuration is mechanically unstable, and the system will display a tendency to have different parts shearing at different strain rates. This is similar to what happens for instance in the van der Waals description of the liquid-gas transition when forcing the system to stay in the coexistence region. The system separates in two phases. In the
present case, due to the geometry of the imposed shear, the two phases correspond to two slabs of material, oriented parallel to the externally applied shear, and deforming at different rates.
We will study this phenomenology in a model system of soft disks \cite{footnote1}  in the presence of an aging attracting interaction.

\subsection{The model}

We follow a simulation scheme similar to that used in \cite{irani} (see also \cite{teitel,durian}).
We work with a system of 3000 particles of radius that are taken from a uniform distribution between 0.8$r_0$ and 1.2$r_0$, with $r_0$ the average value.
 The simulation box is squared, with Lees-Edwards boundary conditions 
along the $y$ direction.\cite{lees}
% that are implemented in the following way. Along $x$ the boundary conditions are strictly periodic, namely
%the velocity of a particle and it image is the same. 
%Along $y$ there is a jump in the $x$ component of the velocity of a particle and its image of value
%$\Delta v_x= \pm \dot\gamma L_y$ (the sign depends whether is the ``upper" or ``lower" image), while the $x$ coordinate
%has a shift $\Delta x=\pm \dot\gamma L_y t$.
 The equations of motion of the particles are the Newton equations implemented in a velocity Verlet form.
The main input to these equation is the value of the force on each particle. This force is the sum of forces between pairs of particles. Two particles $i$ and $j$ with positions ${\bf r}_i$, ${\bf r}_j$ and velocities  
${\bf v}_i$, ${\bf v}_j$ experience a force that is conveniently divided in two terms. There is an interaction force ${\bf f}^{int}_{i,j}$ depending only on positions, and a dissipative force ${\bf f}^{diss}_{i,j}$ depending on velocities and positions. ${\bf f}^{int}_{i,j}$ is calculated from an interaction potential $V^{int}$ that has the following form

\begin{eqnarray}
V^{int}=\frac{\epsilon}{r_0^2}(d_{ij}-R_i-R_j)^2\Theta(d_{ij}-R_i-R_j)+\nonumber\\
+G(\tau)V^{ag}(d_{ij})
\end{eqnarray}
The first term is the elastic potential between particles. Note that the presence of the Heaviside $\Theta$ function  makes this term be purely repulsive. Here $d_{ij}$ is the distance between particles $i$ and $j$ with radius $R_i$ and $R_j$, and $\epsilon$ sets the scale of energy. We will use non-dimensional units
by taking $\epsilon\equiv 1$, $r_0\equiv 1$, and the mass of the particles $m\equiv 1$.
 The second term is the aging potential. It is given in terms of the function $V^{ag}$ depending on the distance between particles. In essence, we intend to model an attractive well between particles of depth $V^0$. However for a molecular dynamics implementation a transition in which the force is continuous is desirable, so we make the simplest polynomial interpolation with this property, which leads to the following concrete form
\begin{eqnarray}
V^{ag}(d_{ij})&=&0 {~~~\mbox{if}~~~} d_{ij}>R_i+R_j\nonumber\\
V^{ag}(d_{ij})&=&-V^0 {~~~\mbox{if}~~~} d_{ij}<(R_i+R_j)\alpha\label{v0}\\
V^{ag}(d_{ij})&=&-V_0\left( \frac{1}{2}+\frac 34 u-\frac 14 u^3\right ) \mbox {otherwise}\nonumber
\end{eqnarray}
with 
\begin{equation}
u\equiv \frac{(1+\alpha)(R_i+R_j)-2d_{ij}}{(1-\alpha)(R_i+R_j)}
\end{equation}
In this way, $V^{ag}$ goes between 0 to $-V^0$ when the distance between particles goes from $(R_i+R_j)$ to $(R_i+R_j)\alpha$ ($\alpha<1$), and the force originated in the potential $V^{ag}$ is continuous everywhere.

Finally, $G(\tau)$ is the {\em aging function}. It depends on the time $\tau$ since the particles have become closer than the distance $R_i+R_j$.  $G(\tau)$ must an increasing function of $\tau$, yet at this point its precise form is undefined. Having in mind the typical logarithmic dependence that aging phenomena display in many different situations,  we have chosen a similar form here. In concrete, we take
\begin{equation}
G(\tau)=\log(1+\tau/30)
\label{G}
\end{equation}
%$V_0=\times 10^{-4}/6 [*]$), $\alpha=0.99$
%In Section \ref{bounded} we will analyze what the consequences of having a different form of the $G(\tau)$ function are.

The dissipative force ${\bf f}^{diss}$ is responsible to remove the energy that is incorporated into the system by the external shearing. In order not to break Galilean invariance this force has to be expressed in terms of local velocity differences between particles. In one of its simplest possible forms we take this force between particles $i$ and $j$ to be given by \cite{teitel,durian}
\begin{equation}
{\bf f}^{diss}_{i,j}=C({\bf v}_i-{\bf v}_j)\Theta(d_{ij}-R_i-R_j)
\end{equation}
The $\Theta$ function ensures that this force acts only if the distance $d_{ij}$ between particles is smaller than the sum of their radii $R_i$ and $R_j$. $C$ is a constant parameter that has to be taken sufficiently large, so kinetic energy does not accumulate in the system during the simulation, while at the same time sufficiently small to avoid that the friction force perturbs too much the Newtonian dynamics of the particles. We got a good compromise with a value $C=0.3$,
 which will be the value used in all results to be presented. We verified that with this parameters the system is in the overdamped regime by doing some simulations with $C=1$ and not observing qualitative changes in the dynamics\cite{irani,teitel}.

In the simulations to be presented, the system is equilibrated at an external pressure value of 0.2
and sheared at a fixed external value $\dot\gamma$ in the $x$ direction. The main output is the stress $\sigma_{xy}$, noted $\sigma$ for simplicity, that is calculated  from the $x$ component of the interaction force between particles through the formula\cite{lees}

\begin{equation}
\sigma=\frac 1{L_xL_y}\sum_{i,j} f^{int}_{i,j}|_x (y_j-y_i)
\end{equation}

\subsection{Results. Shear bands}

\begin{figure}
\includegraphics[width=9cm,clip=true]{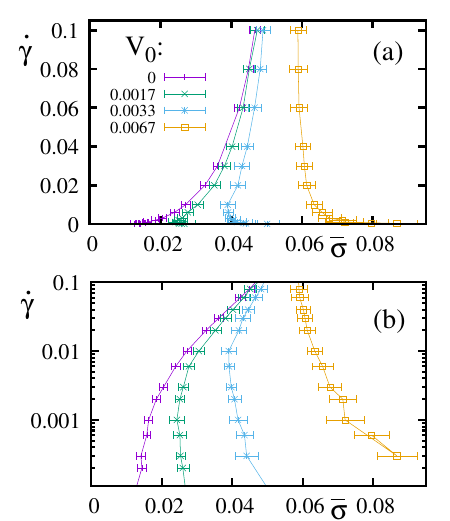}
\caption{Strain rate in linear (a) and logarithmic scale (b) vs average stress in a system of disks without ($V_0=0$), or with ($V_0\ne0$) contact aging, $\alpha=0.99$. The bars indicate the typical fluctuations of the instantaneous values\cite{footnote2}. 
%For $V_0=0$ the expected form of the flow curve $\dot\gamma\sim (\sigma-\sigma_c)^{3/2}$ (\cite{jagla_pt,ferrero-jagla}) is indicated.
In the presence of aging, note the reentrance of the curve at low values of strain rate, and the stronger fluctuations of $\sigma$ in this regime.}
\label{lf0}
\end{figure}

\begin{figure}[h]
\includegraphics[width=8cm,clip=true]{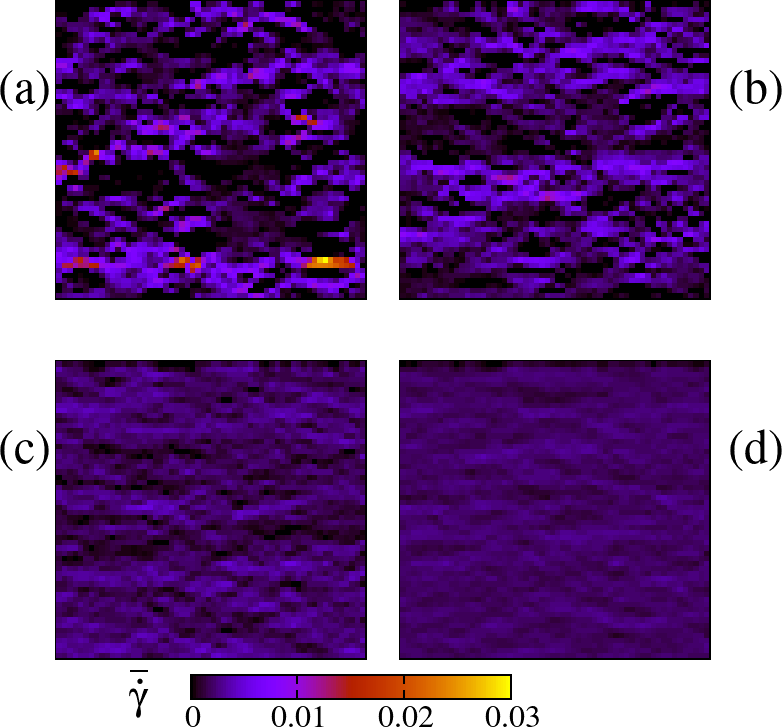}
\caption{Average strain rate at each spatial position in the sample, under an applied strain rate of $\dot\gamma=0.002$,
in absence of aging. This average is calculated in a strain window $\Delta\gamma$ of 
%npasos*.01*.002=0.02 
%npasos=1000,3000,10000,30000
0.02(a), 0,06(b), 0.2(c), and 0.6(d). 
The local average strain rate becomes more uniform as the strain window increases.}
\label{foto1}
\end{figure}

We see results for the flow curve of the system in Fig. \ref{lf0}.
Points in Fig. \ref{lf0} show the temporal averaged values $\overline\sigma$  of the stress, calculated at different values of the applied strain rate $\dot\gamma$. Also, the horizontal bars indicate the typical fluctuations in the instantaneous values of $\sigma$\cite{footnote2}.
Referring first to the case without aging ($V_0=0$ in Eqs. (\ref{v0})), we see 
the standard result for an athermal system of disks (or spheres) under confinement. The flow curve
is monotonous, and there is a critical stress  such that no stationary flow with $\sigma<\sigma_c$ is obtained. 
%Also, we added in Fig. \ref{lf0} a continuous line with the dependence $\dot\gamma\sim (\sigma-\sigma_c)^{3/2}$, which is the expected behavior for this system\cite{jagla_pt,ferrero-jagla}. We see a rather good fitting of the numerical results.
The spatial distribution of deformation in the system is uniform, if averaged on a long period of time. In fact,
in Fig. \ref{foto1} we show  the average strain rate at each spatial position in the sample\cite{footnote3}, under a global applied strain rate of $\dot\gamma=0.002$ in different strain windows $\Delta\gamma$. We see strong inhomogeneities in the deformation in the small strain window cases, but these inhomogeneities average out as the strain window increase, and they decay to zero as the strain window goes to infinity.

When analyzing the results in the presence of aging ($V_0\ne0$), we see a tendency remarkably similar to that observed in the PT model: aging produces a reentrance of the flow curve at low strain rates, and  increase of the temporal fluctuations of the stress $\sigma$.
Yet in the present case, the reentrance of the flow curve has more dramatic effects than in the case of the PT model. In fact, now our system
can be considered as a stack of many layers sliding on top of each other under the action of the externally applied strain rate. We can consider qualitatively that the relative movement of any pair of adjacent layers 
corresponds to a single PT model. Mechanical equilibrium requires that the stress on each layer is the same. If we are in a region of the flow curve with negative derivative, the situation in which the strain rate is the same between each pair of adjacent layers is unstable. In fact, it has long been recognized that such a situation leads to a separation of the system in a region of layers sliding at a larger strain rate that the externally applied values, and another region that remains completely stuck. This is in fact what happens in the present case as we will see now. 

\begin{figure}[h]
\includegraphics[width=8cm,clip=true]{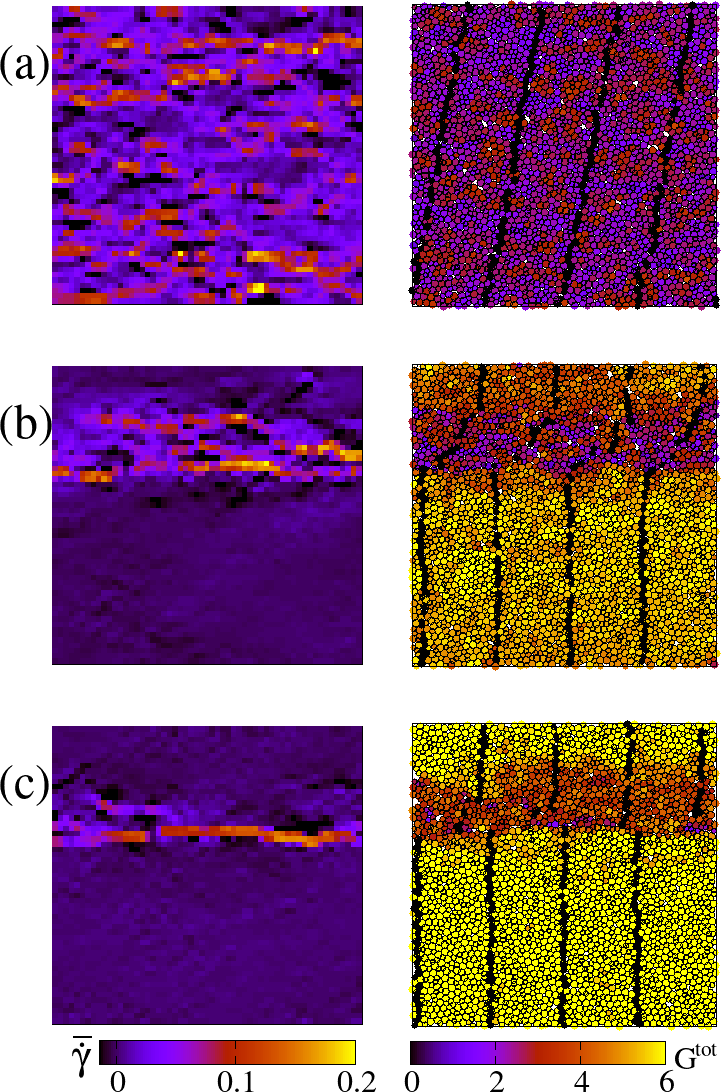}
\caption{Left column: Local strain rate averaged over a strain window of $\Delta \gamma=0.02$ for different
values of applied strain rate:  $\dot\gamma=0.04$ (a), 0.01 (b), and 0.005 (c). In (a) the deformation is uniform across the sample. In (b) and (c) the deformation localizes in a shear band that becomes thinner as $\dot\gamma$ is reduced. Right column: snapshots of the particles in the system. The color code displays the value of the aging function $G^{tot}$ of each particle. We see how particles that are shearing have lower values of $G^{tot}$ than the rest (particles in black are used as a reference. They were vertically aligned at an earlier moment).}
\label{foto2}
\end{figure}

From now on (except when explicitly indicated) we take $V_0=0.01$, $\alpha=0.99$.
Fig. \ref{foto2} (left column) shows the
spatial distribution of deformation in the system in a given time window, at three different values of $\dot\gamma$. In panel (a) the applied strain rate corresponds to a region in which the flow curve has positive derivative, and deformation distributes in a spatially homogeneous way. Panels (b) and (c) correspond to cases in which $\dot\gamma$ is in the region of negatively sloped flow curve. It is clearly seen that in this case there is a {\em shear band} in which the system is deforming, while the rest of the system remains in a frozen configuration\cite{sm}. Moreover, the deformation rate within the shear band takes a rather constant value $\dot\gamma_0$ independently of the globally applied strain rate $\dot\gamma$, in such a way that the shear band width $\delta$ satisfies
\begin{equation}
\frac{\delta}{L_y}=\frac{\dot\gamma}{\dot\gamma_0}
\end{equation}
This expression represents the ``lever rule" similar to that used in the coexistence of first order phase transition. In fact, the reentrant flow curve in the present case is qualitatively similar to the region of negative compressibility of the van der Waals isotherm.% (see Fig. \ref{vdw})

\begin{figure}[h]
\includegraphics[width=8cm,clip=true]{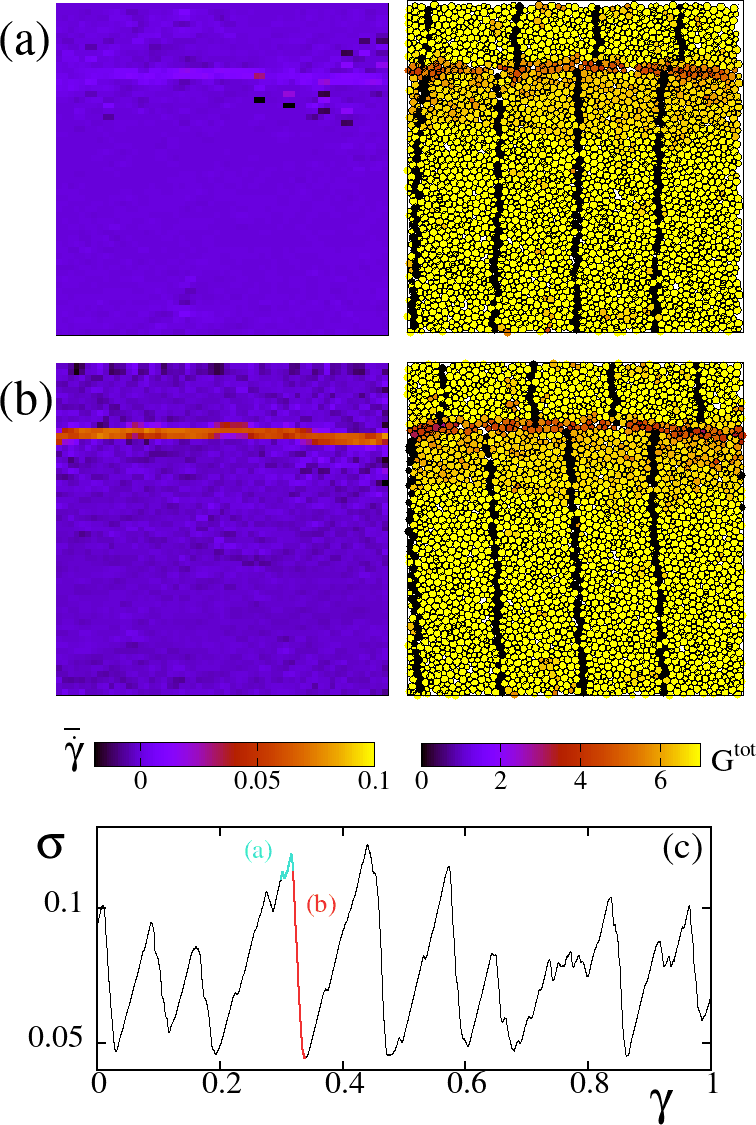}
\caption{(a) and (b): Same  as Fig. \ref{foto2} for deformation at a strain rate
$\dot\gamma=5\times 10^{-4}$. Deformation occurs in a very thin shear band. However, the deformation rate within the shear band is not uniform in time, but acquires a stick-slip behavior, as shown in panel (c). Note in particular the reaccommodation of the lines of particles highlighted in black as one rather large slip event occurs: particles in the shear band experience an abrupt slip, while those in the stuck region have an elastic rebound.}
\label{foto3}
\end{figure}

\begin{figure}[h]
\includegraphics[width=7cm,clip=true]{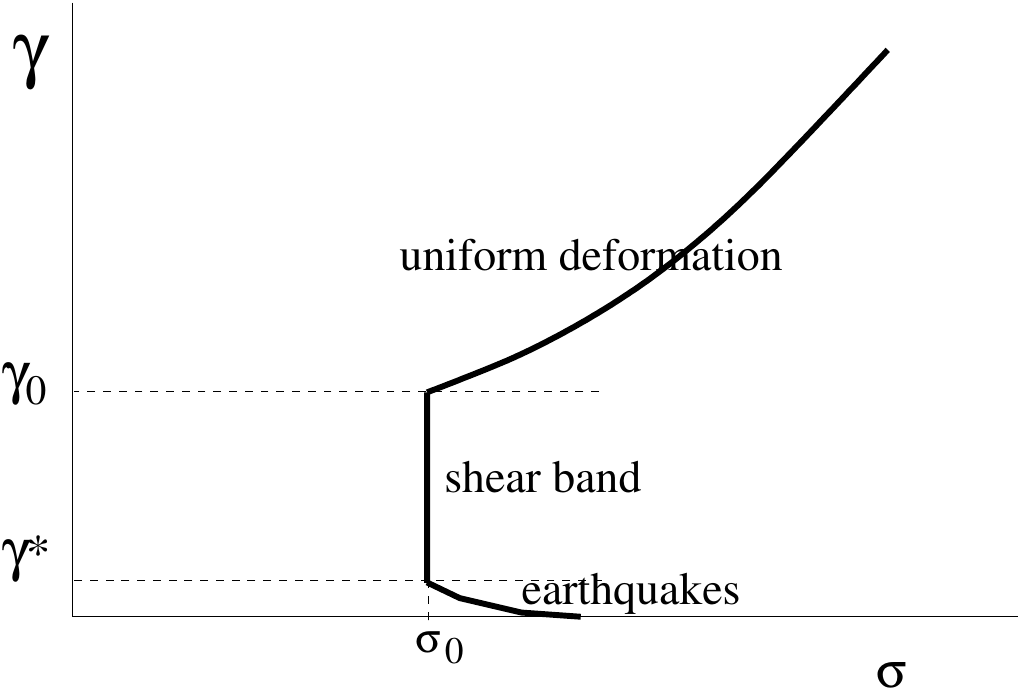}
\caption{Schematic form of the flow curve for the system of soft disks with contact aging under study. The three
qualitatively different deformation regimes are indicated. The value of $\dot\gamma^*$ is size dependant, and is given by $\dot\gamma^*=\dot\gamma_0 \delta^*/L_y$, with $\delta^*$ the minimum thickness of the shear band.
See the text for more details.}
\label{vdw}
\end{figure}

The origin of the shear band is therefore a consequence of the reentrant flow curve, which in turn is due to the existence of the aging mechanism. 
If some part of the system has not yielded for some time, the interaction between particles in it will have developed a larger attractive component making this part more resistant against shear, while particles in regions that are currently yielding have typically a much weaker attractive contribution in the interaction. This produces that regions that are yielding tend to remain in this state, while blocked parts of the system have instead a strong tendency to remain stuck. A direct confirmation of this scenario is obtained by taking the configuration of the system at a given time, and measuring how much aging has built up between particles in such configuration. This can be roughly characterized calculating for each particle $i$ the  value $G^{tot}$ that is obtained averaging the value of $G$ (see Eq. (\ref{G})) between particle $i$ and all overlapping particles $j$ (i.e, $d_{ij}<R_i+R_j$). %, namely
%\begin{equation}
%G^{tot}_i=\sum_jG_{ij}
%\end{equation}
The values of $G^{tot}$ for all particles in the system are shown in the right column in Fig. \ref{foto2}. It is confirmed that particles in the shear band are in fact much less aged than those in the non yielding regions. 

\subsection{Results. Earthquakes}

The idea of a shear band deforming at a constant rate  cannot be sustained if $\dot\gamma$ is too small. In fact, there is a minimum width of a shear band $\delta^*$, set roughly by the size of the particles in the system. If $\dot\gamma$ is so small that $\dot\gamma/\dot\gamma_0 \lesssim \delta^*/L_y$, a uniform deformation rate in the shear band cannot occur. In this regime the deformation at the shear band becomes stick-slip. This is depicted in Fig. \ref{foto3}.  We see that at some time intervals the system deforms almost elastically (a), accumulating stress, while at some other time intervals this stress is released by an abrupt slip in the shear band. This is also the reason why the stress in the system becomes strongly fluctuating in time \cite{sm}.
Fig. \ref{vdw} presents an sketch of the different regimes we have obtained. At the smallest  $\dot\gamma$ the stress departs from the coexistence value $\sigma_0$, becoming larger and also more fluctuating. %This is the ``earthquake regime" of our model.

One of our main points of the present analysis is that the regime of very low strain rate we have just described is  a {\em seismic} regime, with avalanches that can be properly described as {\em earthquakes}.
This claim needs  further justification. In fact, many systems display a response in the form of discrete events generically called avalanches, which cannot necessarily been considered as earthquakes. Although the localization of the deformation in a very thin layer fits with the idea of a seismic fault (with the frozen upper and lower parts of the system representing tectonic plates in relative motion), additional  conditions need to occur in order to speak properly of earthquakes occurring in the system. 
One of these conditions is the fact that the effective friction law between the plates is of the velocity weakening type, i.e., the friction force must decrease as a function of the relative velocity.\cite{scholz} This is precisely what relaxation brings about in our model. The reentrant flow curve is just another way of describing a velocity weakening behavior. 
Actually, it is precisely this behavior that localizes the deformation in a very thin shear band,
generating a ``fault" in which earthquakes can occur. \cite{jagla_2023}

\begin{figure}[h]
\includegraphics[width=8cm,clip=true]{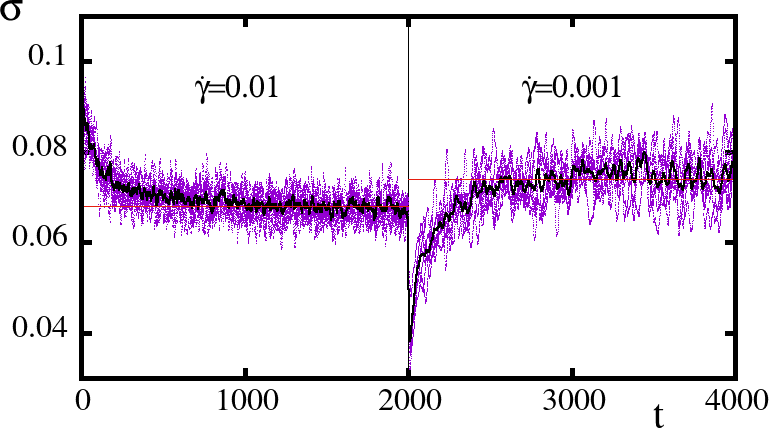}
\caption{Stress in the system under a periodic variation of the strain rate between the two values indicated. The velocity weakening, and
the characteristic overshoot and undershoot at the transitions, typical of the rate-and-state phenomenology\cite{marone} are clearly visible. For this figure we use $V_0=0.0067$, see Eq. (\ref{v0}).}
\label{v1_v2}
\end{figure}

The velocity weakening behavior of seismic faults tells about the behavior under a stationary condition of constant strain rate. There are however also systematic behavior associated to a number of different protocols. One possibility is to vary the applied strain rate in time. The phenomenology is usually well captured by the rate-and-state equations. \cite{scholz,dietrich,ruina,marone}. A full analysis of the behavior of the present model under non-stationary external condition is out of the scope of this presentation, but we want to indicate that in fact the behavior of the model is qualitatively similar to the predictions of rate-and-state equations, and also to the phenomenology of different rocks in the laboratory \cite{marone}. In Fig. \ref{v1_v2} we show the stress in the system when the strain rate is abruptly varied between two different values (namely $\dot\gamma=0.01$ and $\dot\gamma=0.001$). The process is repeated seven times, and the signal is superimposed. The thick black line is the average value. In addition to the velocity weakening effect (that is visible in the larger (smaller) asymptotic value for the smaller (larger) strain rate), we also see the characteristic overshoot and undershoot at the transitions, typical of the rate-and-state phenomenology\cite{marone}.

A second very important piece of phenomenology related to earthquakes is that they generate aftershocks (ASs). 
In practical terms, ASs manifest as an over abundance of quakes occurring after some particularly big event.
It is no the purpose of this work to investigate the detailed statistics of ASs in the present model.
But it is of key importance to show that ASs exist. To address this issue, it is better to describe ASs in a slightly different way. An alternative characterization of ASs is that they are the quakes that occur after the driving on the system is stopped. This points to the fact that ASs are associated to some internal dynamics of the system beyond the one provided by the external loading. In systems with no such internal dynamics all activity ceases (probably after some short inertial relaxation) if external driving is stopped. However, internal dynamics such as the aging mechanism we are considering here can produce instabilities that trigger new quakes even if external driving has completely stopped. The purpose of the following discussion is to show that such instabilities producing ASs exist in the model we are analyzing. 

\begin{figure}[h]
\includegraphics[width=8cm,clip=true]{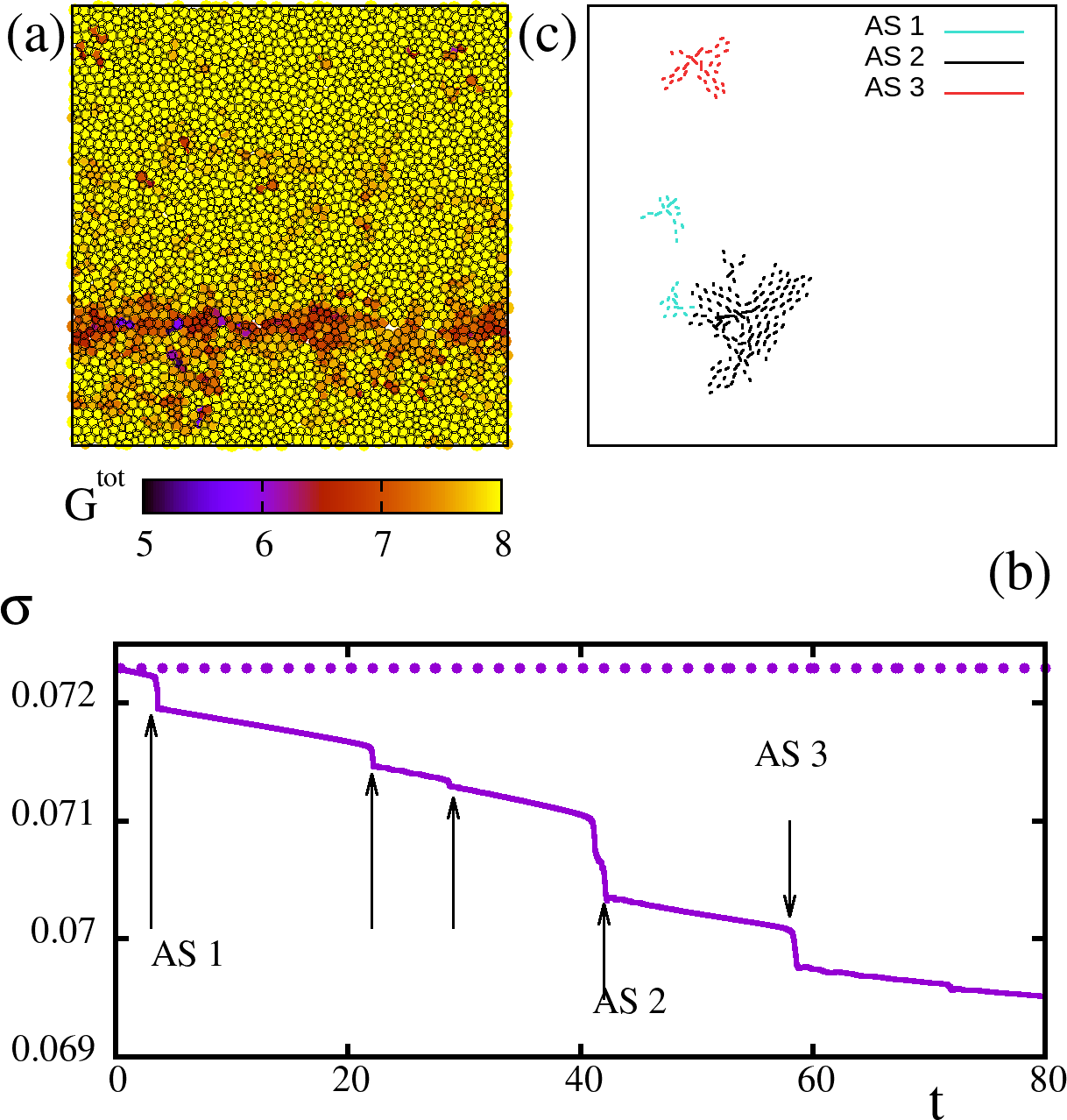}
\caption{(a) The particles in the starting configuration for the AS analysis. The color code represents the value of the aging function $G^{tot}$ of each particle. It is apparent that some particles  of the sample are much less aged than the rest. In fact, it was along this ``fault" that
the system was yielding in the presence of the externally applied strain rate. (b) Stress as a function of time after driving was stopped. Abrupt events (ASs) can be identified, and are indicated by arrows. 
(c) Particle displacements associated to three particular aftershocks (displacement were amplified by a factor of 8 for visualization purposes. Also, only particles experiencing displacement larger than some threshold are shown).}
\label{aftersh}
\end{figure}

\begin{figure}[h]
\includegraphics[width=8cm,clip=true]{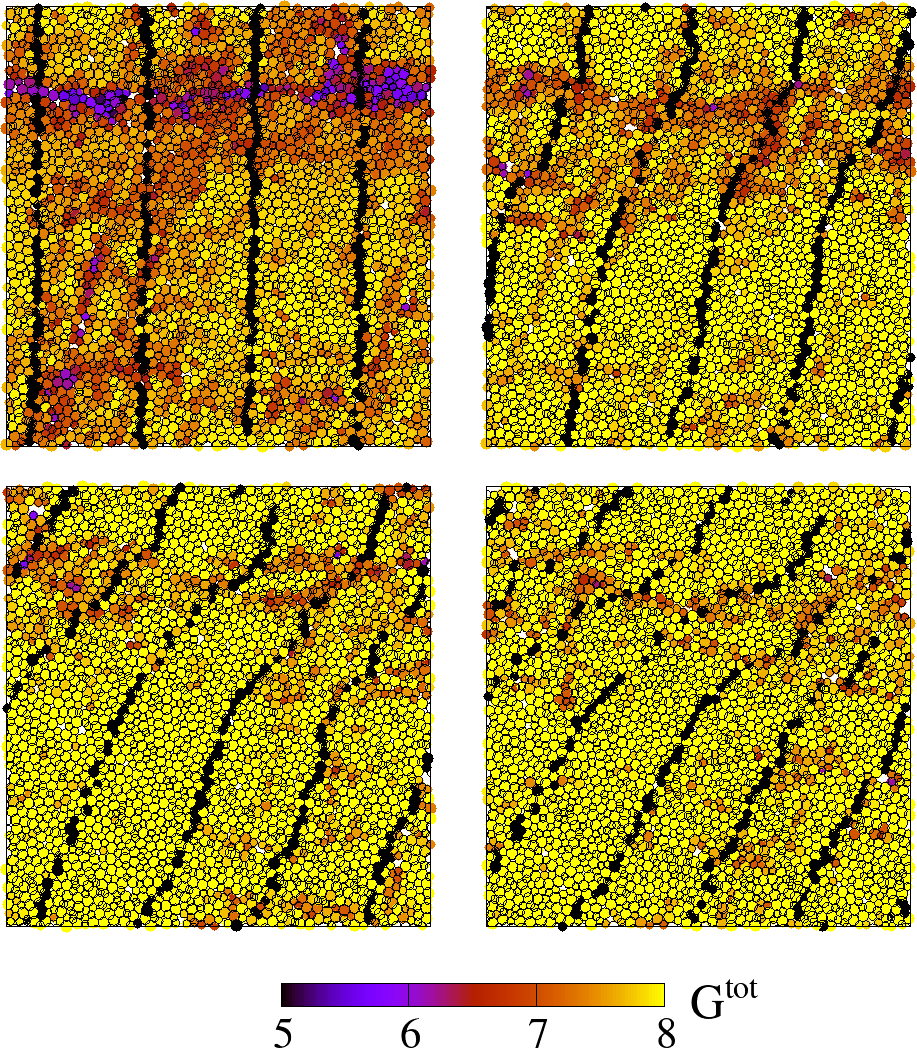}
\caption{Snapshots in a system evolving with $\dot\gamma=5\times 10^{-5}$ (temporal sequence from left to right, and top down).
Colors represent the value of the $G^{tot}$ function of each particle. In the initial configuration there is a 
detectable fault, originated in a previous evolution of the system with a larger $\dot\gamma$. Notice how the fault ``heals", and deformation again occurs uniformly across the sample (as inferred from the black particles), as it happens for the largest values of $\dot\gamma$\cite{sm}.
}
\label{foto3_tanh}
\end{figure}

We use a system with parameters as before, except for the value of $\alpha$, that we take now as $\alpha=0.8$ \cite{cambio_de_param}.
The numerical protocol is straightforward. A starting configuration is obtained after driving the system at a low strain rate, and when there is a clear separation into a frozen region and a very thin shear band where all deformation is taking place. At this point external driving 
is stopped. This defines the initial configuration 
(Fig. \ref{aftersh}(a)). From this, the time evolution of the system is followed. 
The main variable that gives information of the time evolution is the time dependence of the stress in the system. This is plotted in Fig. \ref{aftersh}(b).
The time evolution of stress can be qualitatively described as consistent of two processes. One of them is a quasi-logarithmic decrease of stress with time. This occurs simply because the attractive interaction between neighbor particles increases logarithmically with time, and therefore the measured stress in the system decreases with the same logarithmic law. It  means that this continuous decrease of stress is not associated to any structural change in the configuration of the particles. However,
on top of this continuous decrease there are abrupt jumps, indicated by the arrows. These are the ASs that are originated in mechanical instabilities caused by the aging process. In fact, if aging is also suppressed at the time where external driving is stopped, the stress remains in its original value, and ASs do not occur (dotted line).
% (Fig. \ref{aftersh}(b), dotted line). 
The kind of structural rearrangement that occurs at ASs is depicted in Fig. \ref{aftersh}(c) for three particular ASs. There we show the displacement of particles around the abrupt jump in the stress (magnified by a factor of 8, for better visualization). 

The existence of ASs in the present model is a strong indication that the kind of avalanches that are observed, and the strong stick-slip behavior of the stress in the system can be properly characterized as a ``seismic" phenomenology. Notice that in \cite{weiss} a connection between the characteristics of slips inside a granular shear band
and the properties of earthquakes has recently been made. 

%\begin{figure}[h]
%\includegraphics[width=7cm,clip=true]{sigma_de_t.eps}
%\caption{}
%\label{sigma}
%\end{figure}

\subsubsection{The fate of earthquakes at extremely low strain rates}

The phenomenology discussed in the previous section is appealing as it in fact reproduces many features associated to earthquakes. Coming back to Fig. \ref{foto3}, the punctuated dynamic activity systematically occurring at a quasi-linear location (a quasi-two-dimensional fault for a more realistic 3D system) is a manifestation of a seismic fault in the system. At times in which the system is not yielding (as for instance that of Fig. \ref{foto3}(a)), what keeps track of where the seismic fault is, is the degree of aging of different parts of the sample. In fact, although strain rate is nearly uniform (Fig. \ref{foto3}(a) left picture) there is clear evidence of where the fault is located 
by looking at the aging function (Fig. \ref{foto3}(a) right picture). In this sense, the particles within the fault are a sort of {\em gouge material} as typically referred to in the seismological context. 
However, we may wonder if the strain rate can be so low, that the fault completely heals between two consecutive earthquakes. If this is the case, the next quake will not have any preference to appear at the same spatial position of previous ones.

We have observed in fact that this occurs in the present model if $\dot\gamma$ is reduced sufficiently. 
Using $\dot\gamma=5\times 10^{-5}$ we observe (Fig. \ref{foto3_tanh}) that individual earthquakes typical of this regime do not appear concentrated in a seismic fault, but uniformly distributed across the sample (as observation of the black tracer particles indicate). This occurs because 
no enduring fault remains in the system in the time scale between quakes, and each successive event occurs almost spatially randomly.
We point out that this delocalization phenomenon at very low strain rate has been reported in a mesoscopic model of a yield stress material with aging\cite{jagla_2010}, and also for a material with negative Poisson ratio, in the absence of aging.\cite{poisson}

\section{Summary and conclusions}

In this work we have presented numerical simulations of a model granular material under shear, described as a collection of soft disks, in the presence of an interparticle attraction with a temporal dependence appropriate to describe aging processes in the model. We have shown how the inclusion of the aging attraction transforms a monotonous flow curve with a yield stress, into one with a reentrance, thus allowing the existence of parameter regions in which a stationary shear bands exists in the model. Beyond this general characterization,
the main outcome of this work is to present and characterize a model system that connects two well different 
behaviors occurring at large and small strain rate. On one side, at large strain rate we have a system that can be described as a normal yield stress material. Yet at very low strain rates the behavior corresponds to two consolidated 
plates sliding pass each other and defining a kind of
seismic fault. Within this fault we find the characteristic gouge material in the form of particles that are much less consolidated than the rest of the system, and that follow the rate-and-state phenomenology. Also aftershocks, one of the hallmarks of seismic phenomena, were detected in this regime. The crucial ingredient in the model that allows to obtain this phenomenology is the existence of relaxation, or aging in the particle interaction potential. Incidentally, we have also pointed out to a possible regime of extremely slow driving in which the spatial localization of earthquakes at a well defined fault disappears, as the system is able to heal between successive quakes, which now appear in uncorrelated spatial positions.

%We have presented a model of sticky particles with aging that displays a large piece of the phenomenology corresponding to yield stress materials, linking also to seismic phenomenon as the strain rate in the system becomes very low. 
%The crucial ingredient to obtain this behavior is the inclusion of an aging attraction between particles.
Although the aging attraction  was added {\em ad hoc} to obtain the desired features of the model, it needs to be emphasized that it can be qualitatively justified on physical grounds. The existence of interactions that strengthen with time is widely acknowledged within the
seismological context\cite{scholz}, and it has become clear that they have effect also at the level of mesoscopic contacts.\cite{li,feldmann,vorholzer1,vorholzer2,carpick,carpick0}
 One of its clearest manifestations is in the structure of the rate-an-state equations describing friction between solid bodies.\cite{dietrich,ruina} The strengthening in time of the attraction between two solid bodies in contact can be attributed both to an effective increase of the real contact area with time, and/or to an increase of chemical bonding between atoms in the two bodies.\cite{li} In any case, our time increasing attraction between particles is not unphysical if we consider that we are modeling a granular material in which each particle represents a macroscopic grain.

Finally, in addition to its theoretical interest to study shear bands, seismic phenomena, and their connection in a well defined atomistic model, we believe that the phenomenology described here is accessible  to experimental realizations, for instance in star polymer systems where aging effects leading to shear banding have already been discussed\cite{rogers}.

\section{Acknowledgments}

I thank Ezequiel Ferrero for useful comments on the manuscript.


\begin{thebibliography}{1}



\bibitem{coussot}P. Coussot, 
%Yield stress fluid flows: A review of experimental data,
J. Non-Newton Fluid Mech. 211 (2014) 31–49.

\bibitem{bonn} D. Bonn,  M.  M. Denn,  L. Berthier,  T. Divoux,  and S. Manneville, 
%Yield stress materials in soft condensed matter,
Rev. Mod. Phys. 89, 035005 (2017).


\bibitem{ovarlez} G. Ovarlez, S. Cohen-Addad, K. Krishan, J.Goyon, and P.Coussot,
J. Non-Newton Fluid Mech. 193 (2013) 68–79.

\bibitem{ferrero_rmp} A. Nicolas, E. E. Ferrero, K. Martens, and J.-L. Barrat,
%Deformation and flow of amorphous solids: Insights from elastoplastic models, 
Rev. Mod. Phys. 90, 045006 (2018).

\bibitem{barnes}H. A. Barnes, 
%Thixotropy – A Review. 
J. Non-Newtonian Fluid Mech. 70, 1–33 (1997).

\bibitem{tixo} Coussot, P., H. Tabuteau, X. Chateau, L. Tocquer, and
G. Ovarlez (2006), 
%Aging and solid or liquid behavior in pastes,
J. Rheol. 50, 975.


\bibitem{ovarlez2010}G. Ovarlez, K. Krishan, and S. Cohen-Addad, 
Europhys. Lett. 91, 68005 (2010).

\bibitem{divoux}T. Divoux, D. Tamarii, C. Barentin, and S. Manneville, 
Phys. Rev. Lett. 104, 208301 (2010).

\bibitem{becu}L. Bécu, S. Manneville, and A. Colin, 
Phys. Rev. Lett. 96, 138302 (2006).

\bibitem{mg1} C. Tang, H. Peng, Y. Chen, and M. Ferry, Journal of
Applied Physics 120, 235101 (2016).

\bibitem{mg2}A. L. Greer, Y. Q. Cheng, and E. Ma, Mat. Sci. Eng. R, 74,
71 (2013).

\bibitem{mg3}S. Ogata, F. Shimizu, J. Li, M. Wakeda, and Y. Shibutani,
Intermetallics 14, 1033 (2006).

\bibitem{mg4}C. Zhong, H. Zhang, Q. P. Cao, X. D. Wang, D. X.
Zhang, U. Ramamurty,and J. Z. Jiang, Sci. Rep. 6, 30935 (2016).


\bibitem{jagla2007}E. A. Jagla, Phys. Rev. E 76, 046119 (2007).

\bibitem{fielding}S. M. Fielding, M. E. Cates, and P. Sollich, Soft Matter 5, 2378
(2009).

\bibitem{sollich}P. Sollich, Phys. Rev. E 58, 738 (1998).

\bibitem{coussot2010}P. Coussot and G. Ovarlez, Eur. Phys. J. E 33, 183 (2010).

\bibitem{mansard}V. Mansard, A. Colin, P. Chaudhuri, and L. Bocquet, Soft Matter
7, 5524 (2011).

\bibitem{martens}K. Martens, L. Bocquet, and J.-L. Barrat,
 	Soft Matter, 8 (15), 4197 (2012)
 	
\bibitem{vdb}D. Vandembroucq and S. Roux 	
%Mechanical noise dependent aging and shear banding behavior of a mesoscopic model of amorphous plasticity
Phys. Rev. B 84, 134210 (2011).

\bibitem{varnik}F. Varnik, L. Bocquet, J.-L. Barrat, and L. Berthier, 
Phys. Rev. Lett. 90, 095702 (2003).

\bibitem{berthier} L. Berthier, 
J. Phys.: Condens. Matter 15, S933 (2003).


\bibitem{irani}E. Irani, P. Chaudhuri, and C. Heussinger,
%Impact of Attractive Interactions on the Rheology of Dense Athermal Particles
Phys. Rev. Lett. 112, 188303 (2014)

\bibitem{chauduri}P. Chaudhuri, L. Berthier, and L. Bocquet, 
Phys. Rev. E 85, 021503 (2012).

\bibitem{ragouilliaux}Ragouilliaux, A., G. Ovarlez, N. Shahidzadeh-Bonn, B. Herzhaft, T.
Palermo, and P. Coussot, 2007, Phys. Rev. E 76, 051408.

\bibitem{fall}Fall, A., J. Paredes, and D. Bonn, 2010, Phys. Rev. Lett. 105, 225502.

\bibitem{paredes}Paredes, J., N. Shahidzadeh-Bonn, and D. Bonn, 2011, J. Phys.
Condens. Matter 23, 284116.

\bibitem{dietrich}J. H. Dieterich,  
%Modeling of rock friction: 1. Experimental results and constitutive equations, 
J. Geophys. Res., 84, 2161–2168, (1979). 

\bibitem{ruina}A. Ruina, 
%Slip instability and state variable friction laws, 
J. Geophys. Res., 88, ,359–370(1983). 

\bibitem{scholz}C. H. Scholz, {\it The Mechanics of Earthquakes and Faulting}, 
(Cambridge University Press, Cambridge, England, 2002).

\bibitem{li}Z. Li and I. Szlufarska,
%Chemical Creep and Its Eﬀect on Contact Aging,
ACS Materials Lett.  4, 1368-1373 (2022).

\bibitem{feldmann}M. Feldmann, D. Dietzel, H. Fuchs, and A. Schirmeisen,
%Influence of Contact Aging on Nanoparticle Friction Kinetics,
Phys. Rev Lett. PRL 112, 155503 (2014).

\bibitem{vorholzer1}M. Vorholzer , D. Dietzel, E. Cihan, and A. Schirmeisen,
%Shear-assisted contact aging of single-asperity nanojunctions,
Phys. Rev. B PRL 105, 195401 (2022).

\bibitem{vorholzer2}M. Vorholzer, J. G. Vilhena, R. Perez, E. Gnecco, D. Dietzel, and A. Schirmeisen,
%Temperature Activates Contact Aging in Silica Nanocontacts,
Phys. Rev. X 9, 041045 (2019).

\bibitem{pt1} L. Prandtl, 
%Ein Gedankenmodell zur kinetischen Theorie der festen K¨orper, 
Z. Angew. Math. Mech. 8, 85 (1928).

\bibitem{pt2} G. A. Tomlinson, 
%A molecular theory of friction, 
Philos. Mag. 7, 905 (1929).

\bibitem{pt3} V. L. Popov and J. A. T. Gray, 
%Prandtl-Tomlinson model: History and applications in friction, plasticity, and nanotechnologies, 
ZAMM. Z. Angew. Math. Mech. 92, 683
(2012).

\bibitem{carpick}
K. Tian,D. L. Goldsby, and R. W. Carpick,
%Rate and State Friction Relation for Nanoscale Contacts: Thermally Activated Prandtl-Tomlinson Model with Chemical Aging
Phys. Rev. Lett. 120, 186101 (2018).

\bibitem{footnote1} Although our primary interest is in three dimensional systems, for clarity of the presentation we focus here in a two dimensional case. We have verified that the qualitative findings on which we focus are reproduced in three dimensional systems as well.


\bibitem{teitel} D. Vagberg,  P. Olsson, and S. Teitel,
%Universality of Jamming Criticality in Overdamped Shear-Driven Frictionless Disks,
Phys. Rev. Lett. 113, 148002 (2014) .

\bibitem{durian}D. J. Durian, Phys. Rev. Lett. 75, 4780 (1995).

\bibitem{lees}D. J. Evans and G. P. Morriss, {\em Statistical Mechanics of Nonequilibrium Liquids} (Academic Press, London, 1990).

\bibitem{footnote2}The values of these fluctuations are dependent on the system size, vanishing (except in some particular situations) in the thermodynamic limit. Yet these values provide qualitative information on the system behavior, in particular when comparing their values at different applied strain rates, and in situations with or without aging. 


\bibitem{jagla_pt}E. A. Jagla,
%The Prandtl–Tomlinson model of friction with stochastic driving,
J. Stat. Mech. 013401 (2018).

\bibitem{ferrero-jagla} E. E. Ferrero and  E. A. Jagla,
%Criticality in elastoplastic models of amorphous solids with stress-dependent yielding rates,
Soft Matter, 15, 9041-9055 (2019).

\bibitem{footnote3}To construct this kind of graph we define a 50$\times$50 matrix, and assign each particle to a position $(m,n)$ in the matrix determined by $(m,n)=(\mbox{int} (50x/L_x),\mbox{int} (50y/L_y))$ with $(x,y)$ the spatial coordinates of each particle, and $\mbox{int}()$ the integer part function. Then we assign to the matrix element  $(m,n)$ the numerical value that is obtained  averaging the quantity of interest for particles within that cell. For the concrete case of Fig. \ref{foto2} we first set up a matrix with the values of the $x$ velocity component of particles. At the end we take the difference between this values along the $y$ direction to determine the value of local strain rate that is reported.   

\bibitem{sm}See the Supplementary Material for movies of the dynamics of the system in different parameter regions.
Supplementary files: [xxx].mp4 shows a movie at $\dot\gamma=$[xxx].

\bibitem{jagla_2023}E. A. Jagla,
%Quasistatic deformation of yield stress materials: Homogeneous or localized?
Phys. Rev. E 108, 034123 (2023).

\bibitem{marone}C. Marone, 
%Laboratory-derived friction laws and their application to seismic faulting, 
Annu. Rev. Earth Planet Sci. 26, 643 (1998).

\bibitem{cambio_de_param} Aging produces changes in the force between particles that are at distances between $\alpha(R_i+R_j)$ and $(R_i+R_j)$. If $\alpha$ is too close to one, there are only very few particles that detect the aging process, and the number of ASs is very small in our small samples.
Therefore, for the results in Fig. \ref{aftersh} we have changed the value of $\alpha$, and use $\alpha=0.8$. We keep the same $V_0=0.01$.


\bibitem{weiss}D. Houdoux, A., Amon, D., Marsan, J. Weiss, and J. Crassous,
%Micro-slips in an experimental granular shear band replicate the spatiotemporal characteristics of natural earthquakes, 
Commun Earth Environ 2, 90 (2021).

\bibitem{jagla_2010}E. A. Jagla,
%Shear band dynamics from a mesoscopic modeling of plasticity
J. Stat. Mech. P12025 (2010).

\bibitem{poisson} E. A. Jagla,
%Discontinuous yielding transition of amorphous materials with low bulk modulus,
J. Stat. Mech. 123201 (2021).

\bibitem{carpick0}Q. Li, T. E. Tullis, D. Goldsby, and R. W. Carpick, 
%Frictional ageing from interfacial bonding and the origins of rate and state friction, 
Nature (London) 480, 233 (2011).

\bibitem{rogers}S. A. Rogers, D. Vlassopoulos, and P. T. Callaghan
%Aging, Yielding, and Shear Banding in Soft Colloidal Glasses
Phys. Rev. Lett. 100, 128304 (2008).



\end{thebibliography}
\end{document}